\documentclass[sigconf]{acmart}

\usepackage{booktabs} 
\usepackage{graphicx}
\usepackage{adjustbox}
\usepackage{makecell}
\usepackage{array}
\usepackage{ctable}
\usepackage{float}

\begin{document}

\copyrightyear{2018}

\acmYear{2018}

\setcopyright{acmlicensed}

\acmConference[ICCSP 2018]{2018 the 2nd International Conference on Cryptography, Security and Privacy}{March 16--18, 2018}{Guiyang, China}

\acmBooktitle{ICCSP 2018: 2018 the 2nd International Conference on Cryptography, Security and Privacy, March 16--18, 2018, Guiyang, China}

\acmPrice{15.00}

\acmDOI{10.1145/3199478.3199490}

\acmISBN{978-1-4503-6361-7/18/03}

\title{Data-Driven Threat Hunting Using Sysmon}


\author{Vasileios Mavroeidis}

\affiliation{%
  \institution{University of Oslo}
  \city{Norway} 
}
\email{vasileim@ifi.uio.no}

\author{Audun J{\o}sang}
\affiliation{%
  \institution{University of Oslo}
  \city{Norway} 
}
\email{josang@ifi.uio.no}

\begin{abstract}

Threat actors can be persistent, motivated and agile, and leverage a diversified and extensive set of tactics and techniques to attain their goals. In response to that, defenders establish threat intelligence programs to stay threat-informed and lower risk. Actionable threat intelligence is integrated into security information and event management systems (SIEM) or is accessed via more dedicated tools like threat intelligence platforms. A threat intelligence platform gives access to contextual threat information by aggregating, processing, correlating, and analyzing real-time data and information from multiple sources, and in many cases, it provides centralized analysis and reporting of an organization's security events. Sysmon logs is a data source that has received considerable attention for endpoint visibility. Approaches for threat detection using Sysmon have been proposed, mainly focusing on search engine technologies like NoSQL database systems. This paper demonstrates one of the many use cases of Sysmon and cyber threat intelligence. In particular, we present a threat assessment system that relies on a cyber threat intelligence ontology to automatically classify executed software into different threat levels by analyzing Sysmon log streams. The presented system and approach augments cyber defensive capabilities through situational awareness, prediction, and automated courses of action. 

\end{abstract}

%
%



\setcounter{footnote}{0}
\maketitle

\section{Introduction}

Utilizing threat intelligence has become a priority in cybersecurity operations as a way to prevent an attack or decrease the time needed to discover and respond to an attack. In addition, cyber-attacks are increasingly sophisticated, posing significant challenges for organizations that must defend their data and systems from capable threat actors. Threat actors can be persistent, motivated, and agile, and they use multiple tactics, techniques, and procedures to disrupt the confidentially, integrity and availability of systems and data. Given the risks of the present cyber threat landscape, it is essential for organizations to focus on utilizing cyber threat intelligence and participate in threat information sharing to improve their security posture. In previous work, \cite{mavroeidis2017cti}, we discussed the importance of having access to cyber threat intelligence for increased situational awareness and presented the Cyber Threat Intelligence model that enables cyber defenders to explore their threat intelligence capability and understand their position against the ever-changing cyber threat landscape. Furthermore, in the same work, we commented on the importance of developing a multi-layered comprehensive cyber threat intelligence ontology for improving the threat detection, prioritization, and response capabilities of organizations. The results of \cite{mavroeidis2017cti} indicated that little emphasis had been given to developing a comprehensive cyber threat intelligence ontology, although some holistic initiatives toward that goal existed \cite{iannacone2015developing,syed2016uco,ucobarnum}.

Threat detection and analysis requires aggregating logs into a centralized system known as security information and event management (SIEM). A SIEM collects logs by deploying multiple collection agents that gather security-related events from endpoints, servers, and other security systems and appliances to perform analysis and detect unwanted behavior. In particular, one resource that has received attention for endpoint visibility is Sysmon, a Windows system service and device driver that monitors and logs system activity of Windows workstations. Proposed approaches for threat detection using Sysmon mainly focus on search engines (NoSQL database systems) or graph databases. Without any relevant academic publication, a comprehensive list of related works can be found on GitHub\footnote{\url{https://github.com/MHaggis/sysmon-dfir}}.

The contribution of this paper is twofold. First, we present a comprehensive Cyber Threat Intelligence Ontology (CTIO), based on the CTI model from \cite{mavroeidis2017cti}, and second, we introduce a system for software threat assessment that utilizes CTIO for analyzing Sysmon logs and classifying executed software instances into different threat levels (high, medium, low, and unknown), augmenting defenders cyber defense capabilities through situational awareness, prediction, and automated courses of action.

The rest of the paper is organized as follows. Section 2 explains the importance of utilizing cyber threat intelligence and engaging in information sharing as part of an organization's security operations and discusses how establishing a robust, structured, and expressive cyber threat intelligence knowledge base can strengthen the security posture. Section 3 presents CTIO and elaborates on its composition. Section 4 presents a software threat assessment system that utilizes CTIO and its underlying knowledge base to classify software instances in different threat levels based on the analysis of continuous Sysmon log streams. Section 5 discusses considerations regarding the presented approach. Section 6 concludes the paper.

\section{Threat Intelligence}
Threat intelligence can be described as the aggregation, transformation, analysis, interpretation, and enrichment of threat information to provide the necessary context needed for decision-making \cite{johnson2016guide}. Threat information is any information that can help an organization protect itself against a threat. In a blog post\footnote{\url{http://ryanstillions.blogspot.no/2014/04/the-dml-model\_21.html}}, Ryan Stillions emphasized that security teams of low threat detection maturity and skills would be able to detect attacks in terms of low-level technical observations without necessarily understanding their significance. On the other hand, security teams of high detection maturity and skills are assumed to be able to interpret technical observations in the sense that the type of attack, the attack methods used, the goals, and possibly the identity of the attacker can be determined. 

Threat intelligence sharing allows one organization's detection to become another's prevention by leveraging collective knowledge, experiences, and capabilities to understand better the threats an organization might face. Benefits of threat intelligence sharing include greater insight into cyber threats and enhanced detective and preventive capabilities of an entire community at the strategic, operational, tactical, and technical levels \cite{chismon2015threat}.
Machine-to-machine threat intelligence sharing is facilitated by utilizing machine-readable sharing standards that feed relevant, accurate, timely, and actionable intelligence to threat intelligence platforms. An example is the Structured Threat Information eXpression (STIX) language which is currently the most used standard for sharing structured threat intelligence \cite{sauerwein2017threat}.


\subsection{A Knowledge Base of Threat Intelligence}

A knowledge base is a repository of complex structured and unstructured information that represents facts about the world. A knowledge base can evolve over time and utilize codified logic to infer new facts or highlight inconsistencies. Ontology is a form of knowledge representation that defines semantic concepts and their relationships to elucidate a domain of interest. The agreed-upon schema and unambiguous concepts of an ontology allow information to be structured and form a knowledge base that is queryable and can support reasoning using formal logic.

In previous work \cite{mavroeidis2017cti}, we argued that a comprehensive ontology for cyber threat intelligence would allow organizations of any size to improve their threat detection, prioritization, and response capabilities. Following up, in this work, we developed the Cyber Threat Intelligence Ontology (CTIO). 

\section{Cyber Threat Intelligence Ontology}
Part of our work was to develop a comprehensive Cyber Threat Intelligence Ontology. To achieve that and for supporting interoperability and making the ingestion of cyber threat intelligence into CTIO the least cumbersome, we mainly utilized and interpreted existing works like cyber threat intelligence relevant taxonomies, vocabularies, knowledge bases, and ontologies widely used by defenders. CTIO represents different information types ranging from low-level technical observables to high-level behavioral characteristics, like facts about threat actors, their motivations, their goals and strategies, specific attack patterns and procedures (TTPs), malware, general tools and infrastructures used in adversarial attacks, indicators of compromise, atomic indicators, targets, software weaknesses and vulnerabilities, and courses of action.

We used the web ontology language (OWL) and followed an agile approach for developing the ontology. CTIO comprises several interconnected sub-ontologies based on existing universally utilized taxonomies, such as the Common Vulnerabilities and Exposures (CVE), National Vulnerability Database (NVD), Common Vulnerability Scoring System (CVSS 2.0), Common Platform Enumeration (CPE), Common Weakness Enumeration (CWE), Common Attack Patterns Enumerations and Characteristics (CAPEC), Threat Agent Library (TAL), Threat Agent Motivation (TAM), Adversarial Tactics, Techniques and Common Knowledge (ATT\&CK), sharing standards like STIX 2.1 and OpenIOC, and domain expertise that allowed us to develop a malware ontology and extend the existing CPE schema (ExtendedCPE) to make it more expressive based on our needs. Figure 1 illustrates the interrelationships between the aforementioned concepts. For further information regarding the taxonomies and sharing standards mentioned above and how they relate to the Cyber Threat Intelligence model, refer to \cite{mavroeidis2017cti}.

\begin{figure*}[!h]
\begin{center}
	\includegraphics[width=\textwidth]{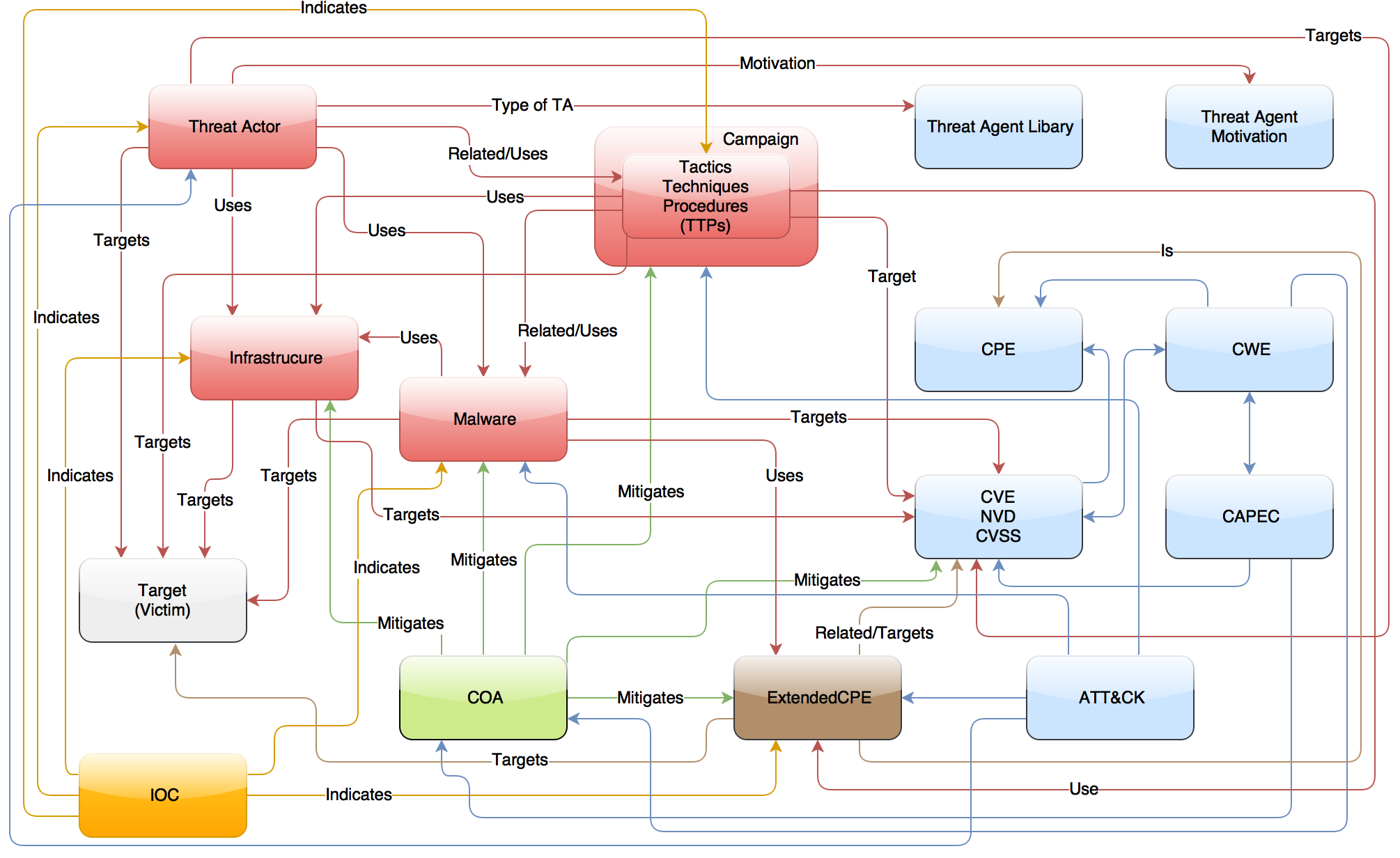}
	\caption{High-Level Relationships of Cyber Threat Intelligence Ontology }
 \end{center}
\end{figure*}

The malware and the ExtendedCPE ontologies are the two major components highly queried in the threat assessment system described in the next section, and they are intended to represent accurate knowledge of malicious and non-malicious software. All the aforementioned ontologies compose a larger unified ontology for representing comprehensive cyber threat intelligence. OWL constructs are used to perform inference and consistency checking over the knowledge base. For example, to classify software as ExtendedCPE, which is a form of whitelist, requires all the classification criteria of CPE to be met and, additionally, to include a process hash followed up by a programmatic verification function confirming that the software classified is deemed non-malicious. It should be mentioned that ExtendedCPE aims to aggregate non-malicious software but includes software with known or unknown vulnerabilities or benign software that has been utilized in adversarial attacks (e.g., command line tools, browsers, vulnerability scanners, network scanners); hence software within the ExtendedCPE subontology can be associated with different threat levels.

The malware ontology was initially developed based on the STIX 2.1 malware object and was later enriched with several other properties to assist our automated software assessment methodology. For example, we included properties that increase the possibility of detecting malware based on Sysmon logs' information, such as hashes and dynamic-link libraries that were loaded during the execution of a malware.

The modularity of CTIO allows utilizing existing ontologies and introducing additional concepts into the main ontology skeleton with minimal integration complexity. Information and documentation about CTIO can be found on GitHub\footnote{\url{https://github.com/Vasileios-Mavroeidis/CTIO}}.

\section{Software Threat Assessment System}
The second contribution of this research work is a system (Figure 2) that utilizes Sysmon logs, cyber threat intelligence, and formal logic to classify executed software on endpoints as of high, medium, low, or unknown threat level based on technical or behavioral characteristics defined in a policy (Table 1) and encoded into the ontology. Thus, organizations can increase their threat awareness capability and partly automate a process for detecting malicious or suspicious software instances on their infrastructure. Also, cyber threat intelligence allows defenders to better understand the threat and how to respond.

The system handles available threat intelligence multi-purposely. Not only can it identify malware based on a principled and systematic analysis but can improve the overall cyber defense operations through increased situational awareness, prediction, and descriptive or machine-executable courses of action.

\begin{table} [!h]
\includegraphics[width=\columnwidth]{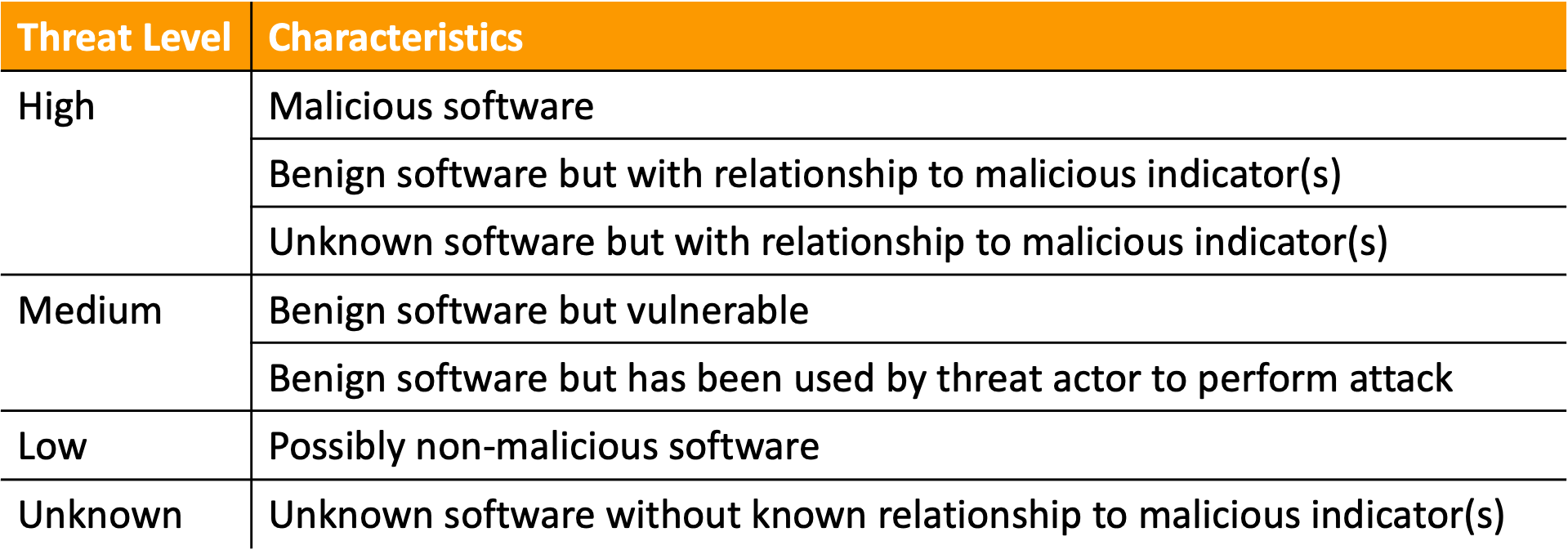}
\caption{Example Threat Level Classification Policy}
\vspace{-1cm}
\end{table}

\textit{Situational awareness}: is achieved through the evidence-based knowledge accumulated within the ontology. A simple observable such as an IP, domain name, hash, or registry key can be part of or related to an indicator of compromise captured within the knowledge base and be queried upon to retrieve more contextual information based on what is known. For example, an identified malicious hash can be pivoted to provide related information about command and control (C2) servers that this malware instance has been observed communicating with, the malware family that belongs to, the campaigns that have utilized this malware instance or another instance of the same family, the threat actor behind the identified campaign and malware, the motivations and goals of the threat actor, as well as the target of the attack such as a specific sector the malware family and the attacker target. When an incident's scope can be determined and taken into account, the response speed and effectiveness increase.

\textit{Prediction}: an organization can introduce an anticipatory threat reduction element into security operations through the increased levels of situational awareness attained from utilizing cyber threat intelligence. For instance, at a more technical level, an unknown executed software that relates to a known malicious property may support revealing an associated malware family. A defender can potentially infer the subsequent steps of a campaign targeting the organization or quickly get an insight into what the attack possibly has caused.

\textit{Course of action}: refers to the steps taken either to prevent an attack or respond to an attack. A course of action within CTIO is described in prose or in a standardized manner that enables real-time automated response actions. Our system utilizes the OASIS Open Command and Control (OpenC2) language \cite{mavroeidis2020nonproprietary}. OpenC2 enables the command and control of cyber defense systems and components in a manner that is agnostic of the underlying utilized products, technologies, transport mechanisms, or other aspects of the implementation. An OpenC2 command comprises an action, a target, an optional actuator that executes the command, and additional arguments that influence how the command is performed.  OpenC2 assumes that an event has been detected, a decision to act has been made, the action is warranted, and the initiator and recipient of the commands are authenticated and authorized \cite{openc2}.\\

Other advantages of the proposed system are the following:

\begin{itemize}
\item Integrating and updating new and existing concepts and threat intelligence is achieved seamlessly or requires minimal modifications due to the system's underlying ontology language technology. Therefore CTIO can be enriched structurally and updated about emerging threats.

\item Sysmon log analysis can help detect threats that could otherwise go undetected by traditional network intrusion detection systems and network firewalls, such as encrypted traffic.

\item The inference capability of ontologies by using logic, the available constructs, and class expressions can derive very expressive knowledge representations increasing data unification and interpretability. For example, a set of rules can classify new malware instances based on the infrastructure type they use, like malware that has used cloud service APIs to exfiltrate data or malware related to establish a botnet and botnet infrastructure. Also, consistency checking is vital to avoid misrepresentation of data.

\item The ontological knowledge base can be searched using granular semantic queries allowing human or machine agents to answer complex questions or to perform threat hunting.
Queries can also be enriched with regular expressions formulating a more signature-based detection method.

\item The proposed system can speed up security operations, improve the detection rate of non-benign software, and add an additional layer of security by automating the investigation process.

\item The cyber threat intelligence ontology can scale, be deployed in a cloud environment, and be maintained by an organization or a threat intelligence community. In our case, CTIO is accessed using rest-style SPARQL queries over HTTPS.
\end{itemize}

\begin{figure}[!h]
\begin{center}
	\includegraphics[width=0.8\columnwidth]{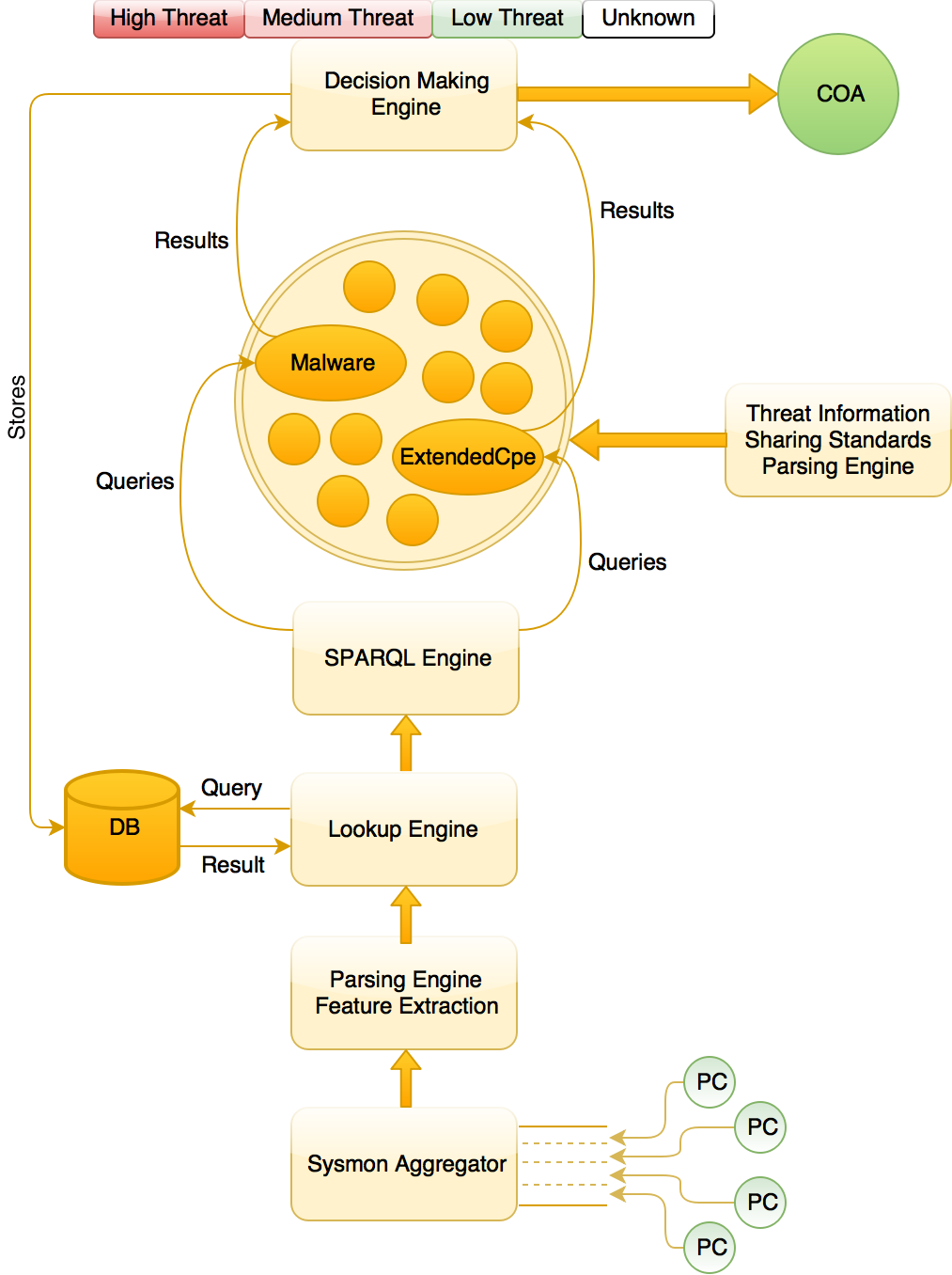}
	\caption{High-Level Architecture of the Software Threat Assessment System}
 \end{center}
\end{figure}

\subsection{Operational Flow of the System}

The system, also presented in Figure 2, \textit{aggregates} Sysmon logs from Windows-based workstations and, using a \textit{parsing engine}, automatically extracts attributes based on each log's Event ID for conducting a threat assessment. For example, a log with Event ID 1 provides detailed information about process creation. Figure 3 presents a simplified Sysmon log with Event ID 1 linked to the WannaCry ransomware attack manifested in May 2017. The parsing engine extracts multiple elements like Event ID, computer name, username, timestamp, process hash, and command lines of both current and parent processes.

\begin{figure}[!h]
\vspace{-0.3cm}
\begin{center}
	\includegraphics[width=\columnwidth]{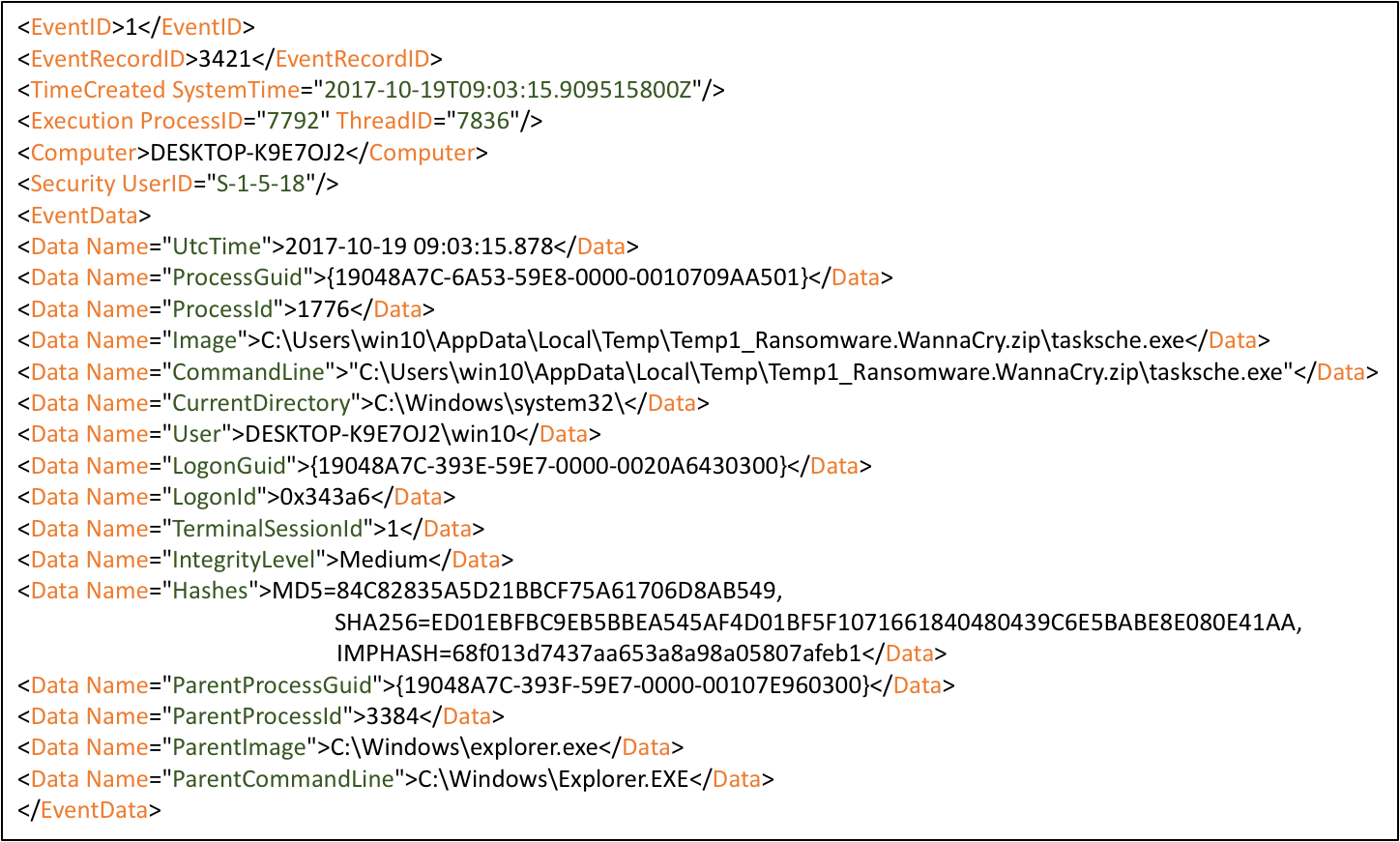}
	\caption{Sysmon Log with Event ID 1 Related to WannaCry Ransomware}
 \end{center}
\end{figure}

Next, a \textit{lookup engine} inspects whether each process is included in an in-house hash whitelist part of the ExtendedCPE component and retrieves the associated threat level. The threat level of a benign process may change based on new information, such as in the case of a new CVE. Also, benign software instances associated with a particular threat level may be further inspected regarding their behavior. For example, a PowerShell instance spawned by a graphical word processing program will raise a case. Further, a downloaded file (Sysmon Event ID 11) by a PowerShell instance spawned by a graphical word processing program will classify the file as a high threat. Such criteria are encapsulated within ontology expressions allowing an inference engine to deduct new information. Also, the relevant Sysmon events that are to be further investigated are mapped, translated to triples, and are included in a dedicated knowledge base. The \textit{lookup engine} inspects whether other extracted element values such as hashes and command lines have been previously queried within a specified time-period and retrieves the relevant information. This tier retains the processing cycles of the \textit{SPARQL engine} low and verifies rapidly benign or malicious software. In the sight of an already classified process, the system pushes the information directly to the \textit{decision-making process engine}, and the appropriate \textit{course of action} is applied or recommended. Element values of unidentified processes become part of SPARQL queries that perform semantic searches upon the CTIO knowledge base and are further transformed into triples for performing reasoning. Based on the derived information and the codified threat classification rules like the ones presented in Table 1, the \textit{decision-making process engine} classifies a process as high, medium, low, or unknown threat level. Processes that have been classified unknown are either considered benign after manual verification or are further investigated in timed intervals by being correlated with new intelligence. Furthermore, the system recommends or executes \textit{courses of action} by referencing a course of action type policy and presents relevant threat intelligence to increase threat awareness.

Figure 4 presents a set of sequential semantic queries based on the WanaCry ransomware process creation Sysmon log presented in Figure 3.
\begin{figure}[!h]
\begin{center}
	\includegraphics[width=\columnwidth]{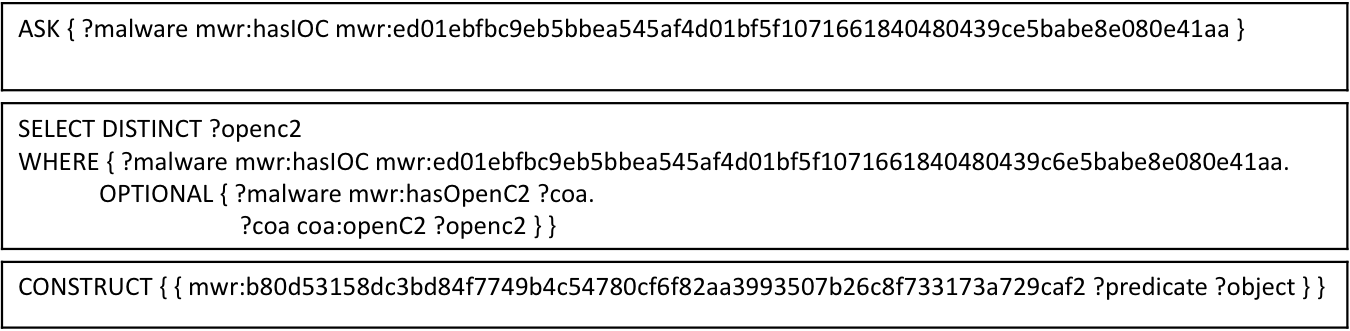}
	\caption{Basic queries in the SPARQL Engine}
 \end{center}
\end{figure}
Given a hash, the first query investigates whether an associated indicator of compromise exists in the knowledge base. Having confirmed an existing indicator of compromise, the system based on the codified inference statements defined in the associated policy has inferred that the process is of high threat. The second query requests a \textit{course of action} to implement and is forwarded to the \textit{decision-making process engine} that, based on a course of action type policy, allows or disallows execution. For instance, in the case of WannaCry, a course of action constitutes allowing traffic passing through a firewall for a specific domain that acts as a kill-switch, blocking C2 communications to specific .onion domains, for externally facing servers and systems that do not use SMB or Windows Network File Sharing capabilities block SMB network traffic, and finally restore infected systems to a previous state. Examples of OpenC2 commands are presented in Figure 5.

\begin{figure}[!h]
\begin{center}
	\includegraphics[width=0.65\columnwidth]{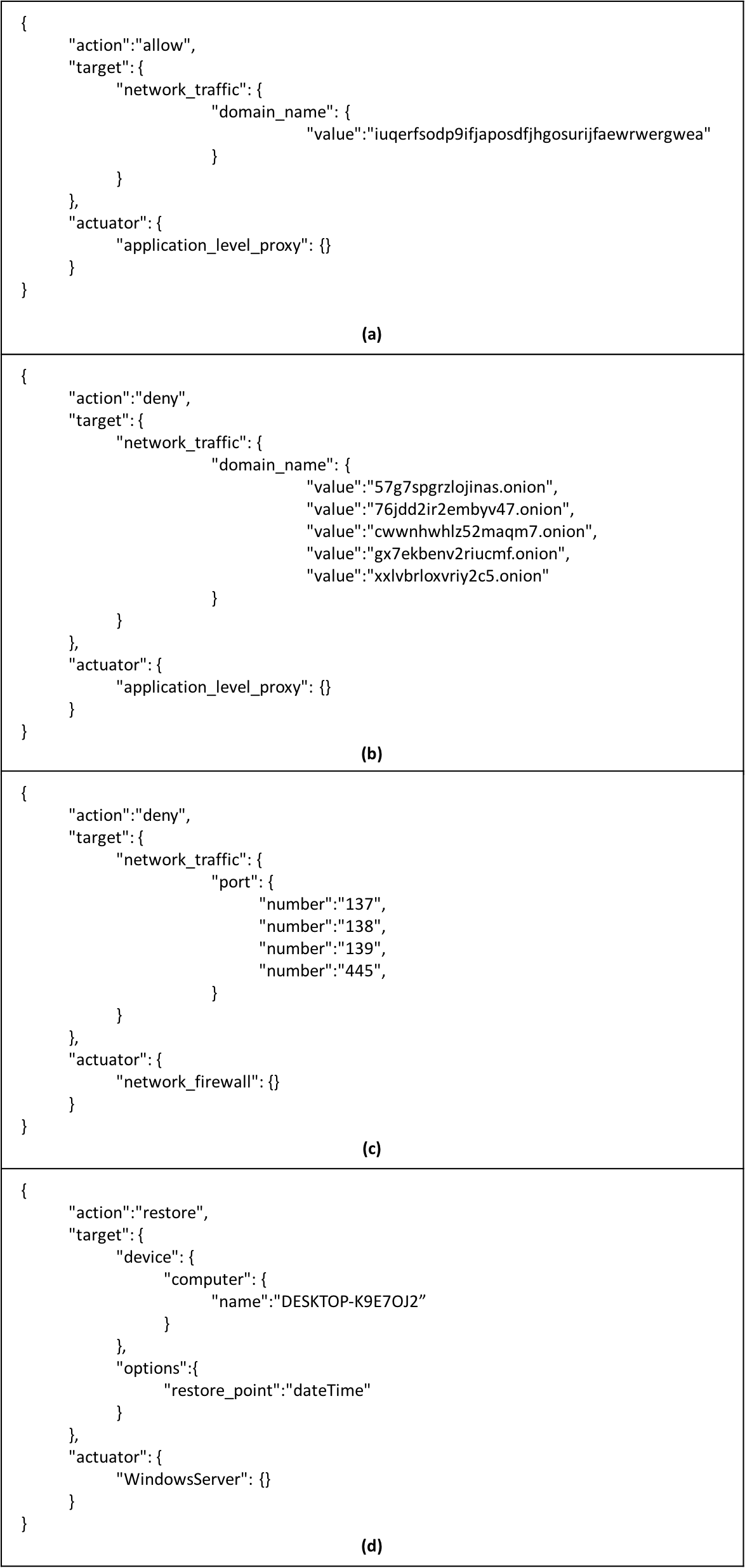}
	\caption{Example OpenC2 Course of Action for WannaCry Ransomware }
 \end{center}
\end{figure}
 
\begin{figure*}[!h]
\begin{center}
	\includegraphics[width=\textwidth]{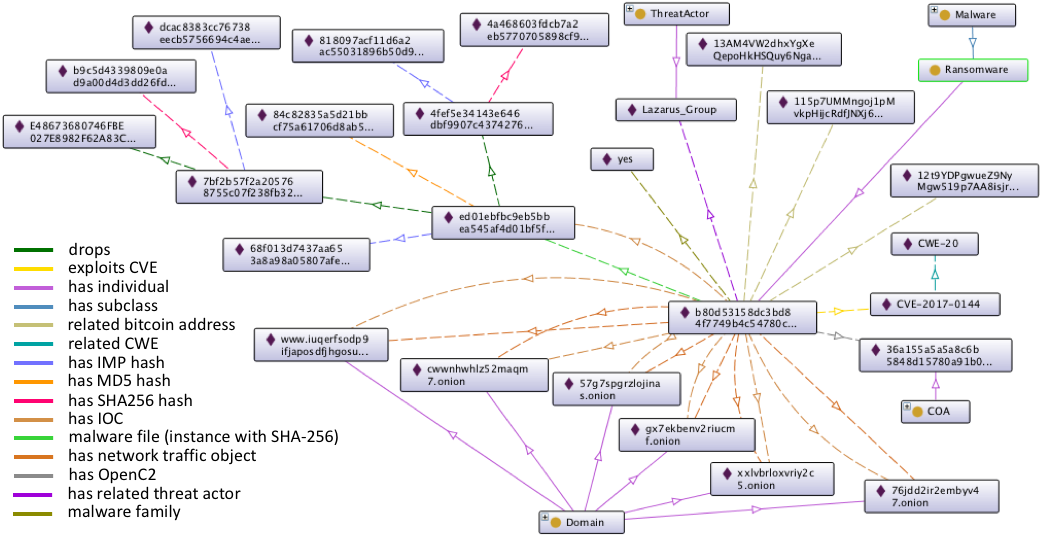}
	\caption{High-Level RDF Graph of WannaCry Ransomware}
 \end{center}
\end{figure*}

Additionally, the system returns a set of RDF triples that comprise the complete known to our organization knowledge regarding the identified threat (third query). Figure 6 presents a high-level threat intelligence graph of the referenced WannaCry ransomware.

\section{Discussion}

This research work presented an approach for software threat assessment that relies on Sysmon logs and a cyber threat intelligence ontology to evaluate and infer the threat level of instantiated software in a system automatically. The general system architecture and the underlying ontology are not restrictive to utilizing Sysmon logs but, in the same way, can utilize a diversified set of log types.

The proposed approach elucidated the benefits derived from utilizing ontology technology and logic for cyber threat intelligence purposes, where manual-based approaches often hinder the complex tasks of correlation, analysis, and inference.  An ontology for cyber threat intelligence comprises multiple concepts describing the who, what, why, when, where, and how of adversarial operations and can be integrated into many different functions of cyber defense such as risk management, threat hunting, incident response, or proactive defense. 

Performing core reasoning tasks and semantic queries on large and complex ontologies are resource and time-intensive. Scaling such ontological systems should be considered by taking into account the size of the knowledge base, the number and complexity of the expressions and rules applied, and the frequency for applying reasoning on new intelligence.

Finally, elevating the standard RDF tabular representation to visualized semantic graphs provides better and easier knowledge exploration and, consequently, conveys key insights more effectively.

\section{Conclusion}
Defenders utilize cyber threat intelligence to make threat-informed decisions. In this research work, we presented a semantic representation of a cyber threat intelligence model \cite{mavroeidis2017cti} using the web ontology language for the purpose of introducing automation in assessing the threat level of instances of executed software on endpoints. Using the reasoning capability of ontologies, we codified statements that can infer the threat level of a process by correlating information derived from Sysmon logs to cyber threat intelligence. In addition, we demonstrated how a standardized language for command and control could activate a rapid threat-informed response.

\begin{acks}
This research was supported by the research project Oslo Analytics (Grant No. 247648) funded by the Research Council of Norway.
\end{acks}

\bibliographystyle{unsrt}
\bibliography{acmart}

\end{document}